\documentclass[a4paper,fleqn]{elsarticle}

\usepackage[numbers]{natbib}

\usepackage{amssymb}
\usepackage{amsmath}
\usepackage{parskip}
\usepackage{multirow}
\usepackage{tabularx}
\usepackage{caption}
\usepackage{subcaption}
\usepackage{optidef}
\usepackage{framed}
\usepackage{ifthen}
\usepackage{lineno}

\begin{document}
\let\WriteBookmarks\relax
\def\floatpagepagefraction{1}
\def\textpagefraction{.001}



\title{Decision making under uncertainty for deploying battery storage as a non-wire alternative in distribution networks}                      



%
\author[1]{Marc Barbar}






\affiliation[1]{organization={Massachusetts Institute of Technology},
    addressline={77 Massachusetts Avenue}, 
    city={Cambridge},
    postcode={02141}, 
    state={MA},
    country={U.S.A.}}

\author[1]{Dharik S. Mallapragada}

\author[1]{Robert Stoner}



\begin{abstract}
The growing demand for electricity in emerging markets and developing economies (EMDE) is causing loading and congestion problems on distribution networks, particularly in urban locations, that adversely impact sustainable development and economic growth. Electric utilities in these economies face unique constraints regarding raising capital required to upgrade their congested networks. Battery storage has emerged as a non-wire alternative (NWA) to feeder upgrades. This article presents a flexible valuation framework for battery storage use in distribution networks and its application in the context of EMDE distribution network planning. We evaluate the value of storage as an NWA using a multi-stage decision making process that combines system optimization with markov-decision processes (MDP) to identify the least-cost network upgrade strategy under demand growth uncertainty. This approach was applied to distribution feeders in Delhi, India, and results highlight the cost-effectiveness of battery storage to manage load growth while deferring network investments. Across the low, medium and high battery storage capital cost projections for 2030, we estimate that 18 to 29 GWh of battery storage capacity could be deployed to defer 11,752 to 15,914 km of medium voltage distribution feeder lines that are loaded at 60\% or more of their ampere capacity in 2030, resulting in 12 to 16\% capital cost savings. Interestingly, lowering storage capital costs does not always lead to increased storage deployment, due to network capacity constraints limiting opportunities for off-peak storage charging.


\end{abstract}




\maketitle

\section{Introduction}

Investments in electrical distribution networks tend to be \textit{lumpy} \citep{UoF2016} since they require large capital commitment initially and  involve significant economies of scale because the assets have long lifetimes (20 to 40 years). Consequently, long-term distribution network planning is often necessary to identify the timing and size of investments needed to meet future demand reliably and cost-effectively while maximizing asset utilization. With declining cost of Li-ion battery energy storage (referred as battery storage here on), there is growing interest to consider its use as a non-wires alternative (NWA) to defer expensive distribution network upgrades and serve rapidly growing peak demand within electricity distribution networks. The modularity of battery storage as well as its flexibility both in terms of location and speed of deployment are in stark contrast to the attributes of conventional network investments and thus represent a potentially valuable option to be considered in network planning. The role for storage as a flexible investment option \cite{Barbar2019} is particularly relevant for loaded urban distribution systems in megacities in fast-growing emerging market and developing economy (EMDE) countries, such as Cairo (Egypt), Delhi (India) and Jakarta (Indonesia), because of several factors \cite{Rao2019, su10051440}. First, many of these cities are experiencing rapid electricity demand growth, due to growing adoption of air conditioners (AC) for space cooling \cite{Barbar2021, McNeil2013} that contributes to network congestion through increasing peak electricity demand \cite{IEAcooling18}. Second, distribution companies in many of EMDE countries are often financially constrained \cite{UN2021} and have to contend with relatively high cost of capital to finance network investments. Third, the premium on land use and geographical constraints in some of these megacities could result in further network investments (reconductoring and upgrading lines) to be operationally challenging or infeasible \cite{WEC2009, WB2014}. Here, we analyze the optimal sizing and placement of battery storage and its economic value as an NWA at the primary feeder level in urban electricity distribution networks of Indian megacities such as Delhi.

India’s electricity demand is projected to more than double by mid-century \cite{brookings}, primarily from increasing electricity use for space cooling in the buildings sector and to a lesser extent, by electrification of transportation \cite{IEAcooling18, IndiaEnergyOutlook}. Much of the growth in energy demand is concentrated in megacities like Delhi, where 55\% of electricity use is residential, which is more than double the national average (24\%) \cite{Barbar2021}. Distribution companies in Delhi are witnessing a level of growth in cooling demand that is capable of overloading the network equipment - for example, feeder data for 2018 indicates that 28\% of feeders were loaded at 60\% or more on an ampere capacity basis. As of 2020, Delhi's peak power demand was 6.7 GW \cite{CEA2018}, and long-term demand projections from our prior work \cite{Barbar2021} suggest wide variation in possible outcomes depending energy needed for space cooling related electricity demand (see Table \ref{table:delhi-ac}). This wide range of possible future outcomes creates significant uncertainty for investment planning in the distribution networks. 

\begin{table}[!h]
\centering
\caption{ Projected peak demand (GW) under the baseline and high-AC-efficiency scenarios assuming stable GDP growth for the city state of Delhi. Further details in \cite{Barbar2021}}
\begin{tabular}{c|cc}
\hline
 & High AC efficiency & Baseline \\
 \hline
2020 & 6.7 & 6.7 \\
2030 & 12.7 & 15.2 \\
2040 & 25 & 36.7 \\
2050 & 34 & 63.8 \\
\hline
\end{tabular}
\label{table:delhi-ac}
\end{table}

Historically, distribution companies have not considered demand uncertainty in their long-term network planning, but have instead resorted to deterministic net present value methodologies \cite{UoF2016, Evans2020}. Demand forecasts enable a comparative assessment of program implementation such as efficiency, policy, and technology under various scenarios. Probabilistic forecasting and flexible planning may be most useful in situations when the magnitude of future outcomes exhibit wide variations, which is the case for the peak electricity demand  projections for a city like Delhi (Table \ref{table:delhi-ac}). Accounting for uncertainty in investment planning may particularly be important when contemplating the use of distributed energy resources (DER) as an alternative to grid expansion, owing to the modularity of DER technologies and their speed of deployment. Until recently, most DER in EMDE countries have been in the form of diesel generators deployed near large commercial and industrial (C\& I) loads\cite{ifc2019}. However, declining costs for Li-ion battery storage\cite{ATB2020} make it a more attractive option. Moreover, battery storage provides the added advantage of not creating local air pollution, a major environmental externality in most EMDE megacities. Furthermore, depending on the energy source used for battery charging, the carbon footprint of energy discharged from battery storage is lower than the diesel generation commonly used in EMDE countries to meet peak demand\cite{6344193}.

Distribution network investment planning has been previously assessed in several studies. For example, in \cite{6450147}, the authors compare the deployment of low-cost diesel generation in rural communities of Latin America to network reinforcement costs while accounting for uncertainty in electricity prices. Another study \cite{KIM2017918} addresses the value of options analysis for DER under future technology cost uncertainty. Authors of another study\cite{DAS2020100482} proposes a flexible investment strategy for renewables incorporation and least-cost system design at transmission level. Such studies that consider uncertainty in electricity system decision making are either concerned with bulk power system design and ignore distribution level planning, or are not concerned with the temporal nature of local DER operations, since their assumed DER are flexible and dispatchable resource (diesel generators, hydro) or energy producing (rooftop solar PV). Moreover, frameworks developed for analyzing role for DER generators are not as informative when evaluating the role for an energy-limited resource such as battery storage that also implicitly couples multiple periods of network operations through its charging-discharging patterns. While regulatory frameworks are being developed for battery storage at distribution level \cite{GRIMM2020114017}, generation design and dispatch models which can assess the different roles of battery storage on a network are generally constrained to deterministic formulations \cite{PILPOLA2019100368, SGOURIDIS20138, Jenkins2017, Knueven2020, SAHLBERG2021100714, OLSSON2021100705} and do not typically consider the impact of long-term uncertainty in different factors influencing planning decisions, such as demand, technology costs and policy evolution. Other studies have considered temporal variability of grid operations and battery storage sizing and operations in the systematic planning of distribution network \cite{9347026, 9088223, batteries6040056}. Multi-stage stochastic programming approaches have also been applied to the problem of distribution network planning, wherein grid operations were modeled using representative periods and investment in energy storage was not considered as a model variable \cite{9582899, 7353222}. In summary, studies considering demand or other types of uncertainties in distribution network planning have typically relied on a limited temporal resolution of system operations and thus may under-value energy storage's ability to alleviate network constraints by shifting generation over time.

Here, we develop a financial valuation framework for battery storage use in distribution network planning that accounts for: a) design and dispatch of battery storage on the electric distribution grid subject to network and operational constraints consider hourly system operations over the year and the b) impact of long-term uncertainty from demand growth on the value and timing of battery storage deployment as a NWA. Our approach is based on combining a linear programming based optimization with a Markov Decision Process (MDP) based simulation that provides a transparent mechanism for evaluating the role for battery storage as a NWA. We illustrate the value of our framework through investigating the potential for battery storage in distribution networks prevalent in Delhi and other megacities in India.  Although our analysis is based on available feeder conditions in Delhi, this approach offers general insights about the conditions under which it is economically viable to defer network investment by deploying battery storage. Moreover, while our method can readily incorporate investment in other types of DERs like rooftop PV, we only consider battery storage NWA for the case study of distribution feeders in India due to its novelty as a technology, modularity and, minimal space footprint, which is an important practical consideration for a highly congested urban megacity. We also ignored diesel generators due to their CO$_2$ and air pollution related externalities. In summary, the key contributions of this paper are:

\begin{enumerate}
    \item The development of an hourly network expansion optimization model at the distribution level that is used to compare the relative cost-effectiveness of battery storage NWA and traditional network upgrades while accounting for independent sizing of battery storage power and energy capacity and operations.
    \item The application of the optimization model in a multi-stage simulation-based environment to evaluate the value of battery storage as an NWA under long-term demand growth uncertainty and various battery storage technology cost scenarios.
    \item Quantifying the potential for using battery storage as an NWA in the distribution networks for four major Indian megacities.
\end{enumerate}

The paper is structured as follows: Section \ref{method} describes the methods, including the key assumptions and structure of the proposed model; Section \ref{param} details the Delhi case study input data we use; Section \ref{results} describes results for Delhi and other megacities in India; in Section \ref{discussion}, we discuss the cost implications for the Indian case study. Finally, Section \ref{conclusion} describes the conclusions, as well as its limitations and potential areas for future work.

\section{Methodology}\label{method}
\begin{figure*}[!ht]
\centering
 \includegraphics[width=12cm,keepaspectratio]{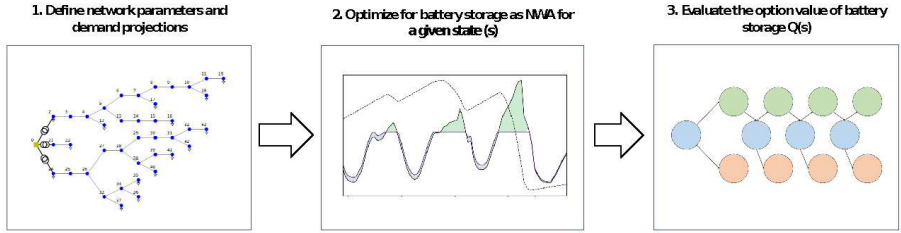}
  \caption{Flowchart showing steps in the flexible valuation framework used for modeling storage as a NWA.}
   \label{fig:summary}
\end{figure*}

The model for evaluating battery storage as a NWA is divided into three categories: financial, technical, and probabilistic. Therefore, our overall approach is divided into three broad steps. Step 1 (section \ref{m1}) presents the valuation criteria for battery storage to be financially feasible as a NWA given network parameters and demand projections as modeled in Fig. \ref{fig:summary} Box 1. Battery storage is utilized as a flexible investment to defer large network upgrades. Thus, battery storage costs and deferred network upgrade costs must be cheaper than traditional network upgrades, otherwise battery storage is not financially feasible. We must also solve for the battery storage system, location, size and operation on the network. In this context, step 2 (Section \ref{m2}) involves the system design optimization with hourly dispatch to size and place battery storage as a NWA on a network. A simulation of various demand growth trajectories will finally inform the network investment planning under demand growth uncertainty. Step 3 (Section \ref{m3}) describes a Markov decision process (MDP) of exploring the various system designs from section \ref{m2} that satisfy the battery storage NWA valuation criteria of section \ref{m1} to identify a least-cost network investment planning framework that adequately considers battery storage. Fig. \ref{fig:summary} highlights the overall flexible valuation framework. The overall objective of the flexible valuation framework is to explore near-term solutions using storage as an asset to minimize the overall capital expenditure on the system by taking into consideration long-term demand growth uncertainty.

\subsection{NWA valuation criteria}\label{m1}
Distribution companies may defer long-term investments by deploying battery storage to meet their short-term peak demand needs and mitigate short-term financial commitments. For a battery storage NWA system to be beneficial, the net present cost of deferring traditional network investment must outweigh the battery storage system cost. This condition is shown in Eq. \ref{eqn:oc}. Given a planning horizon starting in period $y_0$ and ending in $y_n$ where $n$ is the number of years to consider demand growth uncertainty, the valuation criteria considers capital ($I$) at time $t$ and the total fixed ($F$) and variable ($V$) costs during the planning horizon $[y_0,y_n]$. Three cost structures are compared: traditional network upgrades $l$, battery storage $b$ and deferred network upgrades $d$. The right-hand size of line 1 of Eq. \ref{eqn:oc} refers to the costs of traditional network upgrades $l$ over the entire planning horizon $[y_0,y_n]$. Lines 2 and 3 of Eq. \ref{eqn:oc} splits the planning horizon into two subsets: $[y_0,p]$ and $[p,y_n]$. The first subset refers to periods in which battery storage is used as a NWA and network upgrades are deferred until period $p$. The second subset refers to the remainder of the planning horizon starting at $p$ and ending at $y_n$. All three incurred costs ($I$, $F$, $V$) are considered for battery storage and deferred network upgrades during their respective lifetimes. Variable costs for battery storage include charging cost at the available wholesale electricity tariff.

\begin{equation}
\begin{split}
    O_t^s(p)&= C_{l,t}^I+\sum_{t \in [y_0,y_n]} (C_{l,t}^F+C_{l,t}^V)\\
                & - C_{b,t}^I+\sum_{t \in [y_0,p]} (C_{b,t}^F+C_{b,t}^V)\\
                & - C_{d,t}^I+\sum_{t \in [p,y_n]} (C_{d,t}^F+C_{d,t}^V)
\end{split}
\label{eqn:oc}
\end{equation}

For a given demand growth scenario $s$,$O_t^s(p)$ in Eq. \ref{eqn:oc} defines the cost of deferring traditional network investments $l$ by $p$ periods by installing battery storage $b$ at time $t$ and subsequently upgrading the network ($d$) at period $p$. This refers to the option value of battery storage NWA for network deferrals. If $O_t(p) < 0$ then battery storage is a financially feasible NWA for a set of costs, deferral period $p$ and planning horizon.

\subsection{System design optimization}\label{m2}

The system design optimization stage evaluates the cost-optimal location, sizing and dispatch of storage subject to operational constraints. This is achieved by formulating and solving a linear program for capacity expansion and dispatch \cite{gitgenx} of a power system network \cite{gitegret} as described in Appendix \ref{optmodel}. The model objective is to minimize the total system cost which includes annualized resource expansion (generation, storage, networks) and, operational costs as described in Eq. \ref{obj}. The operational constraints are: 1) balance of system at the hourly level (Eq. \ref{dem}), 2) time-dependent capacity constraints for generation resources (Eq. \ref{vre}), 3) battery storage state of charging, energy, power capacity limits and degradation (Eq. \ref{stor1} - \ref{stor4}, Eq. \ref{obj}), 4) generation unit commitment (Eq. \ref{therm1} - \ref{therm1_2}), 5) generation minimum and maximum power (Eq. \ref{therm2}, \ref{therm3}), 6) generation ramping limits (Eq. \ref{therm4}, \ref{therm5}), 7) direct current power flow approximation through line susceptance and voltage deviation (Eq. \ref{net1}), 8) network flow limits (Eq. \ref{net2}, \ref{net3}) and, 9) non-negativity constraints (Eq. \ref{nnc}). 

For the purpose of placing and sizing battery storage on the system in the context of distribution networks in India, we restrict the linear program of Appendix \ref{optmodel} in the following ways: a)we consider storage deployment exclusively at the feeder node of the network, b) we do not conside existing or investment in distributed generation in the network since the overall objective is to relieve congested lines in highly dense urban cities with minimal space for DERs like PV and c) we do not consider the battery storage's ability to inject power upstream since battery storage feed-in tariffs were not yet established in India when this study was carried out \cite{tata_tariff}. We assume that there is enough upstream generation (from the transmission system) to meet the demand on the feeder, hence the maximum available generation capacity is greater than peak demand. Additionally, the minimum upstream generation supply and ramping limits are set to meet the minimum demand and time step change in load. The optimization model has three investment variables, namely storage capacity, storage power, and network line capacity upgrades. 
Non-served energy is included in the objective function to allow for the possibility of feasible solution via load shedding. This is a single-stage optimization so the design of the system is based on the inputted annualized investment costs and demand for a given time period only. Under this formulation, the supply-demand balance of  Eq. \ref{eqn:oc} is enforced while respecting storage capacity constraints, network flow constraints. The optimization identifies the capacity of battery storage to be deployed, only if it is cost-optimal for the current stage and consistent with the valuation criteria defined in section \ref{m1}. In particular, storage dispatch must adhere to network flow constraints both during charging and discharging periods. The model is formulated in Pyomo \cite{hart2017pyomo} and solved using CPLEX \cite{cplex2009v12}. 

\subsection{Simulation}\label{m3}
A Markov decision process (MDP) is a sequential decision problem for an observable and stochastic environment with a Markovian transition model and discounted rewards. It consists of a set of states, a set of actions, a transition model, and a reward function. The sequence of decisions in the distribution network planning problem can be modeled as an MDP given the uncertainty in demand growth. Each state has a value that is calculated using the system design optimization described in Section \ref{m2}. At each state there are two possible actions: traditional network upgrades, battery storage NWA upgrades. The transition matrix $P_{i,j}$ models conditional probabilities of growth in electricity demand from state $i$ to state $j$. Policy iteration is a solution for MDP involving two steps: (1) estimating value function for a given policy, (2) using the estimated value function to find a better policy. Given possible recursion of the two-step process, the value function ($Q$) that is generated under a policy $D$, which maps every state to a decision is expressed in Eqn. \ref{eqn:mdp} where $F(s)$ is the cost resulting from the system design optimization at state $s$ and decision variable $x$, $P(s,s')$ is the probability of transitioning from state $s$ to $s'$ and $\gamma$ is a discount rate. The MDP iterates through various sequences of actions (policies) to find the highest reward value (as per Eq. \ref{eqn:mdp}) given a state transition matrix.

\begin{equation}
    Q^D(s) = F(s)+\gamma\sum_{s'}P(s,s')Q^D(x, s')
    \label{eqn:mdp}
\end{equation}

\begin{figure}[!ht]
\centering
 \includegraphics[width=8cm,height=8cm, keepaspectratio]{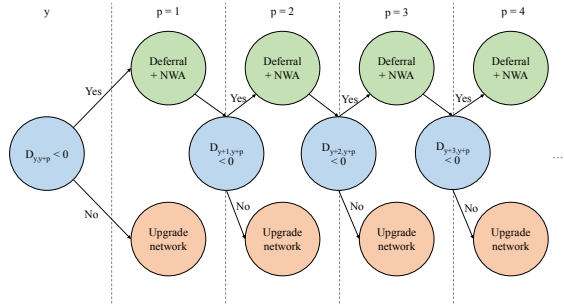}
  \caption{Flowchart of flexible valuation policy iteration.}
   \label{fig:fc}
\end{figure}

To simplify the iterative process of the MDP, we restrict the action of battery storage NWA upgrade to future states in which $O_t^{s'}(p) < 0$, i.e. use of battery storage as NWA is justified (see Eq. \ref{eqn:oc}). Otherwise, the simulation terminates with traditional network upgrade action as seen in Fig. \ref{fig:fc}. Therefore, a decision to take the battery storage NWA action is only possible when $D$ of Eqn. \ref{eqn:ro} is less than 0, which refers to the expected value of option across all states. The MDP can thus be translated to the flexible valuation framework as: $F(s)$ is the total system design cost at stage $s$ (current stage) resulting from the system design optimization of section \ref{m2}, $P(s,s')$ is the probability of transitioning from state $s$ to $s'$, $D$ is the policy option between traditional network upgrade and battery storage NWA, $x$ is the quantity of upgrades (traditional or storage) that is optimized using the system design optimization model described earlier (and available in Appendix \ref{optmodel}), and $Q$ is the new system cost given $x$ at stage $s'$.

\begin{equation}
    D_s(p) = \sum_{s'} P_{s,s'}\cdot O_t^{s'}(p)
    \label{eqn:ro}
\end{equation}

The MDP explores the multi-stage stochastic decision making process by iterating over a chain of policy decisions given the available options at each stage. In other words, this is a discounted sum of real options evaluated in a chain of decisions. At each stage, when the battery storage NWA option is feasible (Eqn. \ref{eqn:oc}) is satisfied), the battery storage system is placed and sized through the system design optimization (see Appendix \ref{optmodel}). This process is repeated for the various trajectories. The resulting system cost of a policy $D$ from $s$ to $s'$ under a transition probability $P(s,s')$ and decision $x$ to add storage or upgrade network given the existing system cost $F$ at stage $s$, populates the MDP Eq. \ref{eqn:mdp}. Note that a policy is only explored when Eq. \ref{eqn:ro} is satisfied, meaning that on expectation it is possible to place storage that satisfies the flexible valuation Eq. \ref{eqn:oc}. If storage is not viable (i.e. Eqn. \ref{eqn:oc} is not satisfied), the policy is therefore to expand the network which ends the MDP iteration. Each case takes less than 1 second to build and solve on a personal computer making it efficient to solve numerous times to get the total annualized investment cost for a stage $s$ and plug it in as the value function $Q(s)$ of the MDP Eq. \ref{eqn:mdp}.

\section{Delhi Case Study Input Data}\label{param}

We demonstrate the value of the flexible valuation framework described above through a case study of network planning for megacities in India, such as Delhi, till 2040 under demand uncertainty.

\subsection{Demand scenarios}
The demand scenarios used as inputs to the network planning problem are adapted from the results of a previously documented demand forecasting model \cite{Barbar2021} that produces hourly resolved electricity demand data at the state-level for various technology and growth scenarios. From the dataset of scenarios developed previously \cite{Barbar2021}, we select three scenarios to define low, medium and high demand outcomes: 1) high AC efficiency coupled with stable GDP growth, 2) baseline AC efficiency and stable GDP growth and 3) high AC efficiency and rapid GDP growth. \footnote{The baseline AC efficiency scenario corresponds to electricity sales projections based on presently available AC units and a high-efficiency scenario that projects adoption of efficient AC units as defined by a recent study \cite{IEAcooling18}. As of 2018, the sales-weighted average Seasonal Energy Efficiency Ratio (SEER) for AC units in India was 3 and the global average was 4 \cite{IEAcooling18}. Under the baseline scenario, the gap between India's SEER and the global average is maintained. Under the high-efficiency scenario, India's SEER rating is projected to reach 8.5 by 2050 \cite{IEAcooling18}}. The high AC efficiency scenario translates into less electricity demand on a distribution network and is therefore considered the \textit{low} growth scenario. The baseline efficiency under stable GDP growth is considered the \textit{mid} growth scenario since it models a business-as-usual outcome. The \textit{high} growth case is selected as rapid GDP growth, meaning strong economic growth and spending power, which leads to higher electricity demand growth but also higher AC efficiency since spending power is higher enabling stronger sales of efficient AC units.

\begin{figure}[!ht]
\centering
 \includegraphics[width=8cm,height=8cm, keepaspectratio]{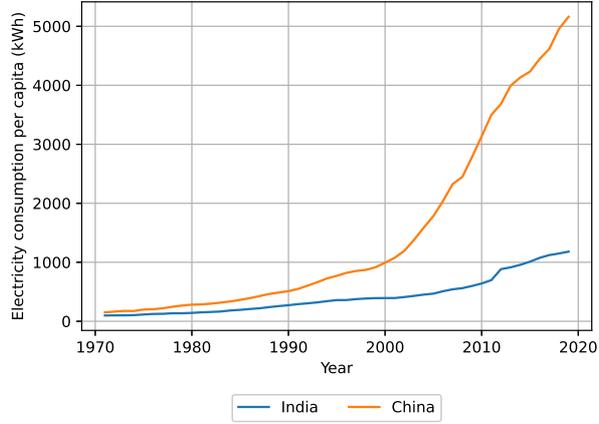}
  \caption{Historical electricity consumption per capita for India and China \cite{WB2021}}
   \label{fig:consum}
\end{figure}

Given the three electricity demand scenarios up to 2050 (low, mid, high), we use Monte Carlo Markov Chain (MCMC) to estimate the stationary distribution of demand growth in India by sampling from electricity demand consumption data from China. While China has achieved faster growth than India in the past four decades, as seen in Fig. \ref{fig:consum}, it is anticipated that India will experience high growth over the coming decades \cite{IndiaEnergyOutlook} . Moreover, Fig. \ref{fig:consum} positions India two decades behind China in electricity consumption per capita as of 2019. India's projected electricity demand, primarily due to space cooling, is strongly compared to China's electricity consumption trend over the past two decades \cite{IEAcooling18}. With India's consumption growth trend following China's, we use the MCMC to produce an estimate distribution of electricity demand growth by sampling from the Chinese electricity consumption data (Fig. \ref{fig:consum}) as the prior information on future electricity consumption in India. 

\begin{figure}[!ht]
\centering
 \includegraphics[width=8cm,height=8cm, keepaspectratio]{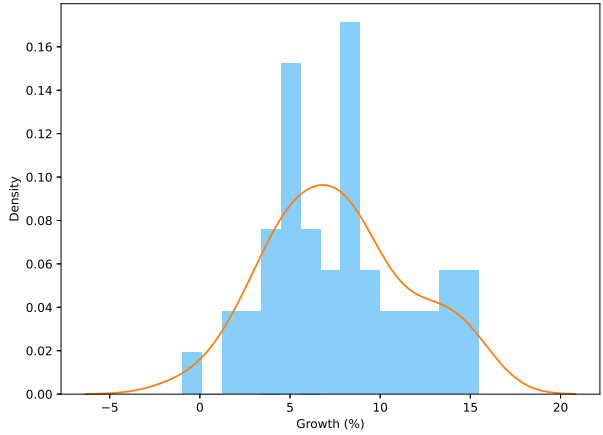}
  \caption{Stationary distribution result of MCMC simulation}
   \label{fig:data}
\end{figure}

MCMC simulations enable estimating a stationary posterior distribution. For every generated random value $x$, a transition kernel is used to assess the parameters of the desired distribution. The transition kernel is split into two steps: a proposal step and an acceptance/rejection step. Given the samples used to approximate the prior distribution, a proposal distribution, and an acceptance criteria, the MCMC iterates until the posterior distribution is stationary. The proposal distribution used in the simulation is a Gompertz distribution \cite{Barbar2021, gompertz}. We define the acceptance criteria as the log error between the prior samples and the proposal distribution. Running the MCMC simulation yields a stationary distribution of demand growth as seen in Fig. \ref{fig:data}.

The result of the MCMC is a continuous distribution of projected electricity demand growth in India. We assume that every projected period of demand growth is an independent event that is sampled from the probability distribution derived by the MCMC. Therefore we construct a transition matrix of low, mid and high growth scenarios as defined in \cite{Barbar2021}. Low growth is assumed to be less than 5\%, high growth is assumed to be larger than 8\%, with mid growth referring to the in-between range. For two independent event with probability distribution, the joint probability distribution that falls within a range of values is a bivariate distribution. The bivariate distributions defined by the demand growth ranges are bucketed into a table. This table serves as the transition matrix in the MDP. Table \ref{table:tm} refers to the expected values of each bivariate distribution defined by the joint probability of column and row events. During the MDP, we sample from this table to simulate a chain of events.

\begin{table}[!ht]
\centering
\caption{Proposed transition Matrix}
\label{table:tm}

\begin{tabularx}{6cm}{c|c|c|c}
\hline

&Low & Mid & High \\
\hline
Low&0.34&0.33&0.33\\
Mid&0.38&0.32&0.3\\
High&0.2&0.8&0\\
\hline
\end{tabularx}

\end{table}

\subsection{Distribution network data}
\begin{figure}[!hb]
\centering
 \includegraphics[width=8cm,height=8cm, keepaspectratio]{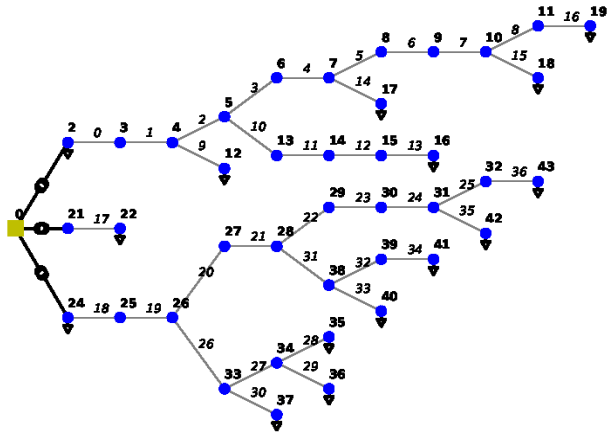}
  \caption{Sample distribution network diagram: three main feeders: commercial (top), industrial (middle), residential (bottom). Bus number in \textbf{bold} and line number in \textit{italic}. Size, length, parameters and topology vary across the library of feeders analyzed \cite{tata}.}
   \label{fig:net}
\end{figure}

We apply the flexible valuation framework method to a benchmark medium voltage distribution network \cite{1709447} and adapt the network equipment --- line, transformer, voltage, current --- to match the data of primary distribution (33 kV and 11 kV) networks operated by Tata Power Delhi Distribution Limited \cite{tata}. As an illustrative example, we model a 1 MW distribution network that is divided into three main feeders leaving the distribution substation: residential, commercial, and industrial as seen in Fig. \ref{fig:net}. Using 2018 loading reports \cite{tata}, the substation is initially loaded at 50\% of the rated 1 MW capacity. We assume a loading limit of 90\% of the rated capacity of the network, in this case, the loading limit is 900 kW.

\subsection{Cost assumptions}

Relevant cost inputs used in the modeling are presented in Table \ref{table:cost}. Storage life is set to 15 years as per other storage assessment reports \cite{ATB2020, MOCI2019, NRELCD2019}. We model battery storage degradation ($C^d$ in Eq. \ref{obj}) as a 1.46\% \cite{LAZARD2020} per annum energy capacity capital expenditure cost premium. Specifically, we assumed costs for battery storage NWA to be the same as transmission-level storage. As discussed later on, higher storage cost assumptions will reduce the option value of battery storage as an NWA. Since storage is an energy-dependent resource as described in the system design optimization (section \ref{m2} and Appendix \ref{optmodel} Eq. \ref{dem}), battery charging ($\psi$) will be from upstream generation which has a variable cost ($C^V$) which varies throughout the day as per Table \ref{table:cost}. Therefore any variable cost incurred to charge the battery storage is associated with the cost of battery storage NWA as per the system design optimization.

\begin{table}[!ht]
\centering
\caption{Input cost assumptions for the model. Sources \cite{ATB2020, NRELCD2019}. Peaks hours are defined to be between 8 PM and 12 AM.}
\begin{tabular}{c|cc}
\hline
&2030 &2040\\
\hline
Energy Cost (USD/kWh) & 168 & 147 \\
Power Cost (USD/kW) & 146 & 128\\
O\&M Cost (USD/kW-yr) & 20 & 18\\
\hline
New line (USD/km) & 350,000 & 350,000\\
Reconductoring (USD/km) & 650,000 & 650,000\\
\hline
Off-peak Tariff (USD/MWh) & 55 & 55\\
Peak Tariff (USD/MWh) & 90 & 90 \\
\hline
Discount rate $\gamma$ (\%) & 9 & 9 \\
\hline
\end{tabular}
\label{table:cost}
\end{table}

\section{Results}\label{results}
We first present the result of a single policy iteration of the MDP for a demand growth trajectory that is sampled from the stationary transition matrix. The resulting model outcome will demonstrate the process through which storage is sized and the policy $D$ is evaluated as per the flowchart of Fig. \ref{fig:fc}. Subsequently, we present the result of the full MDP, aggregated at the city-level for the four Indian megacities,  Delhi, Mumbai, Bengaluru, and Kolkata. The input data that vary by the city are demand projections and distribution network characteristics sourced from elsewhere \cite{Barbar2021, prayas2012}. We evaluate battery storage NWA options for a deferral period of 5 years in the single network case. We further expand the deferral window to 10 years when we evaluate the Indian megacities cases.

\subsection{Single network case --- Delhi}\label{res:delhi}

\begin{figure}[!ht]
\centering
 \includegraphics[width=8cm,height=8cm, keepaspectratio]{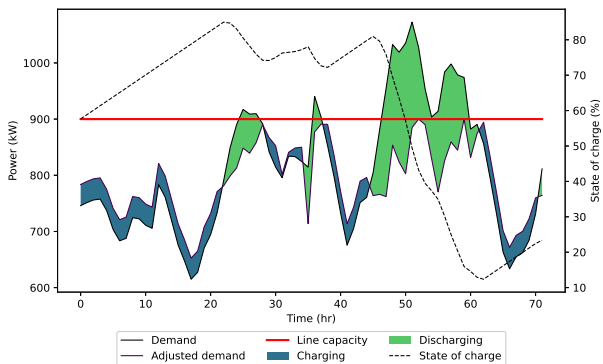}
  \caption{Hourly dispatch of battery storage NWA for one summer week load profile in 2030 for a distribution network from the city of Delhi. The demand scenario is based on the mid demand scenario.}
   \label{fig:batopt}
\end{figure}

We detail the results of storage design on the distribution network for the demand trajectory based on the mid-level demand projection \cite{Barbar2021} in 2030. As previously mentioned in the input data, the initial feeder capacity is 1 MW and it is 50\% loaded in 2018.  The system design optimization yielded a solution of 1.5 MWh battery storage capacity with 380 kW power (5.5 hours duration) on bus 28 of the feeder (node on residential trunk feeder in the network as seen in Fig. \ref{fig:net}). Note that although the system design optimization was constrained to only one location for storage deployment, it is possible to relax the problem to allow for multiple storage systems on the network may be cost-optimal. Fig. \ref{fig:batopt} presents the dispatch of the installed battery storage with respect to the substation's hourly load profile. From the perspective of the substation, the peak demand does not exceed 900 kW (loading capacity of the network) since the battery storage discharge is used to satisfy some of the demand during peak demand periods.

The optimization is repeated for the three possible demand trajectories (low, mid, high). The results are shown in Table \ref{table:bat}. To further illustrate the results of the system design optimization, we solve two variants of the valuation framework: one with only traditional network upgrade and another with battery storage NWA + deferred investments. The results of the annualized investment cost for each demand projection's system design solution are shown in Table \ref{table:net}. We define the annualized investment cost (AIC) as the total cost of an option (capex + opex) with annualization performed using a discount rate of 9\% \cite{rbsa}. Battery storage NWA AIC includes the variable charging cost during off-peak hours. In the case where storage is not included in the optimization as a decision variable (traditional network upgrade only), the flow balance constraints (see Appendix \ref{optmodel}) result in 2,3 and 4 kilometers of line upgrades required under the low, mid and high demand growth scenarios, respectively.

\begin{table}[!ht]
\centering
\caption{Optimized storage sizing in the single network case for Delhi and cost outcomes. Results correspond to three demand growth scenarios (low, mid, high) in 2030 for Delhi using sample network of Fig. \ref{fig:net} and cost assumptions of Table \ref{table:cost}. CAPEX is the annualized capital cost of energy and charge, OPEX is the fixed and variable operation and maintenance costs for one year. VAR is charging cost of the battery storage system given the tariff schedule.}
\label{table:bat}
\begin{tabular}{l|ccc}
\hline
 &Low&Mid&High\\
 \hline
 Power (kW)& 300& 380&420\\
 Capacity (kWh)&1,200&1,520&1,680\\
 \hline
 CAPEX (USD) & 8,358 & 10,586 & 11,701 \\
 OPEX (USD) & 6,000 & 7,600 & 8,400 \\
 VAR (USD) & 4,950 & 8,778 &12,474 \\
\hline
\end{tabular}
\end{table}

\begin{table}[!ht]
\centering
\caption{Annualized investment cost (AIC) results for traditional network upgrades and battery storage NWA with deferred network upgrades for a deferral period $p = 5$. Results for case of single network in Delhi and three demand growth scenarios (low, mid, high) for 2030.}
\label{table:net}

\begin{tabular}{ccc}
\hline
\multicolumn{3}{c}{Traditional network upgrade AIC (USD)}\\
\hline
Low & Mid & High \\
14,673 & 22,009 & 29,345 \\
\hline
\hline
\multicolumn{3}{c}{battery storage NWA AIC (USD)}\\
\hline
Low & Mid & High \\
12,969 & 19,453 & 29,937 \\
\hline
\end{tabular}
\end{table}

Using the transition matrix produced by the MCMC (Table \ref{table:tm}), we run a large number of iterations of the MDP until the chain of policy is stationary. To further elaborate on the results, Table \ref{table:rov} highlights the various option costs $O$ and thus possible policy decisions $D$ (from Eq. \ref{eqn:oc} and \ref{eqn:ro}). Note that the results of the high scenario are negative, that is also reflected in the AIC comparison of Table \ref{table:net} where the battery storage NWA AIC under high growth is more expensive than the traditional network upgrade AIC in 2030. In Table \ref{table:rov}, we note that the option value is positive for the low and mid demand growth scenarios, implying that deferral of network upgrades by battery storage NWA is cheaper than traditional network upgrades on expectation (Table \ref{table:net}). On the other hand, given high demand growth, it is cheaper to upgrade the network immediately. The flexibility of battery storage NWA allows the utility to adopt a "wait and see" strategy and benefit from lower than anticipated growth to defer upgrades. For this reason, the option value of low projection is the highest, and the high one is the lowest (negative) as seen in Table \ref{table:rov}.

\begin{table}[!ht]
\centering
\caption{MDP simulation results for the case of single network in Delhi across all transitions in Table \ref{table:tm} for year 2030. Results are annualized USD values. Negative values indicate that battery storage NWA is more expensive than traditional upgrades and vice versa.}
\label{table:rov}
\begin{tabular}{l|ccc|c}
\hline
& \multicolumn{3}{c|}{battery storage NWA AIC}& Policy value\\
\cline{2-4}
& Low & Mid & High & \\
\hline
Low & (1,361) & 36,328& 74,018 & 35,692\\
Mid &36,328  & 15,864 & 5,632 & 20,679 \\
High & (32,057)  & 5,632  &43,322 & (24,519) \\
\hline
\end{tabular}
\end{table}

\begin{figure}[!ht]
\centering
 \includegraphics[width=8cm,height=8cm, keepaspectratio]{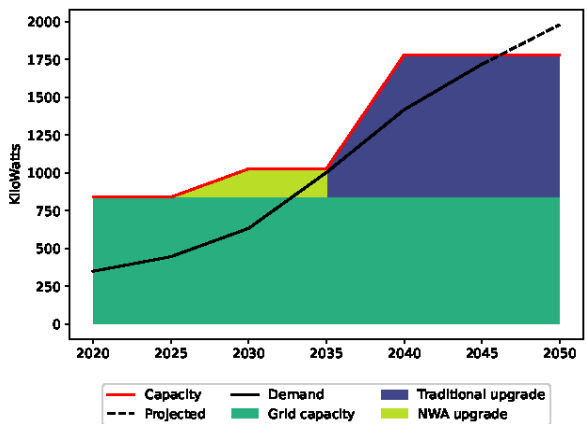}
  \caption{Simulation of investment outcomes for single network case in Delhi, shown in Fig. \ref{fig:fc}, with simulated demand trajectory from the transition matrix. $y=2020$, $Y=20$, four timesteps $p$ of 5 years intervals as per the input projected data \cite{Barbar2021}.}
   \label{fig:sim}
\end{figure}

Iterating through the various policies $D$ at various stages and exploring the available flexible options $O$ yields a stationary chain of decisions given various demand growth trajectories. Fig. \ref{fig:sim} illustrates the final result of the full MDP given the expected demand growth trajectory that is simulated from the transition matrix. In the earlier periods of the modeling, there is a large uncertainty on the level of demand in the final period (2050). Therefore, when a policy needs to be chosen (2030) due to network overload, the MDP elects to place battery storage as opposed to upgrading the network. In later periods, the uncertainty of demand in 2050 yields an early retirement of the battery storage system and a traditional network upgrade that satisfies the demand growth until the end of the simulation period (2045).

The early retirement of storage is taken into consideration when it is installed in 2030 since the policy would not have been an option if Eqn. \ref{eqn:oc} was not satisfied. The option value differs based on technology costs, transition probabilities, demand growth, and desired deferral periods. Evidently, a higher cost of storage yields results that favor traditional network upgrades. On the other hand, lower network reconductoring and new line costs will also favor traditional network upgrades. For the particular case of Delhi, storage costs were chosen from mid-range projections of their respective periods \cite{ATB2020} and the network upgrade costs were collected from benchmark surveys as well as historic upgrade costs of local distribution utility \cite{tata, NRELCD2019}.

\subsection{City-wide network simulation for four Indian megacities}

We apply the flexible valuation framework to estimate battery storage NWA across select megacities in India (Bengaluru, Delhi, Kolkata, Mumbai) that collectively accounted for 52 TWh of electricity consumption in 2018 with an estimated 72,763 circuit kilometers of distribution lines at 33 and 11 kV serving dense urban areas \cite{CEA18, TPDDL} (see Table \ref{table:mega}). For each city, we apply the above-described method to study nine representative feeders identified. The representative feeders were identified based on clustering of the library of urban feeders (and their respective hourly demand profiles) for Delhi provided by TPDDL \cite{tata}. Each representative feeder is characterized by:

\begin{enumerate}
    \item Loading percentage varying from 40 to 80\% of ampere capacity \cite{tata}
    \item Represented demand, defined as the hourly load profile modeled on the feeder, which varies by megacity according to available survey data \cite{prayas2012}
    \item Serviced demand, defined as the total annual demand (MWh) that the distribution network feeders service with the same loading percentage
    \item Serviced circuit kilometers, defined as the total circuit kilometers (km) that are at the corresponding loading percentage
\end{enumerate}

\begin{figure*}[!ht]
\centering
 \includegraphics[width=12cm,height=12cm, keepaspectratio]{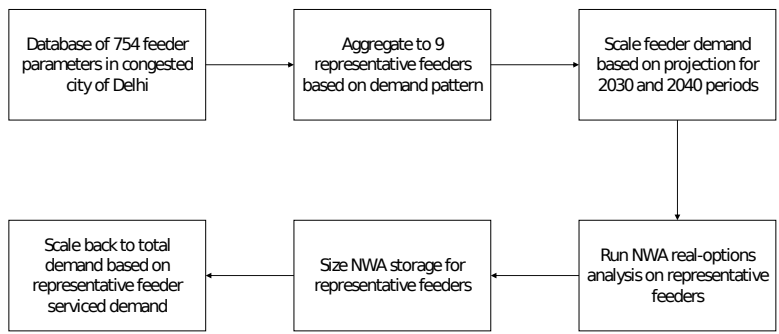}
  \caption{Approach for computing megacity-level battery storage as a NWA in distribution networks of Indian megacities.}
   \label{fig:8}
\end{figure*}

We assume a similar distribution in feeder loading for the other megacities as indicated by the data for Delhi. We estimate the circuit kilometers that each of the nine feeders represents in every megacity based on the respective data on the serviced demand and the calculated ratio of serviced demand to circuit kilometers available for these feeders in Delhi. The flexible valuation framework is applied on all the representative feeders of the four selected megacities by using the appropriate demand projections in 2030 and 2040 \cite{Barbar2021} and the same transition matrix (Table \ref{table:tm}). Network investment costs are calculated based on the circuit kilometer length of each representative feeder in each megacity. The resulting battery storage NWA capacity for the representative feeders is scaled using the calculated ratio of each feeder's serviced demand to represented demand. To explore the impact of storage technology cost on the flexible valuation framework, we consider alternative cost trajectories for battery storage \cite{ATB2020} as well as calculate the cost of storage where battery storage NWA is no longer competitive with traditional network upgrades (referred as the break-even storage capital cost). Finally, as noted earlier, since electricity is mostly contracted in Delhi and other cities in India, we do not consider the value of energy arbitrage where battery storage NWA can sell excess energy back to the grid since as of 2020, there was no tariff schedule to accommodate such transactions.

\begin{table*}[!ht]
\centering
\caption{Flexible valuation framework results for megacity-level battery storage NWA analysis under mid-range storage capital cost projections}
\begin{tabular}{l|c|cccc}
\hline
\multicolumn{1}{l|}{} & Year & Bengaluru & Delhi & Kolkata & Mumbai \\
\hline
\multicolumn{1}{l|}{Demand (TWh)} & 2018 & 10 & 23 & 4 & 15 \\
\hline
\multirow{2}{*}{battery storage NWA (GWh)} & 2030 & 3 & 14 & 1 & 11 \\
 & 2040 & 15 & 50 & 35 & 40 \\
 \hline
\multirow{2}{*}{Overloaded lines (km)} & 2030 & 1,265 & 6,093 & 792 & 12,224 \\
 & 2040 & 1,467 & 7,070 & 919 & 14,184 \\
 \hline
\end{tabular}
\label{table:mega}
\end{table*}

Under the mid-growth scenario, we estimate that 20,373 km of 72,763 km of the four megacities' distribution networks will be overloaded by 2030. Under the same demand growth projection, an additional 23,640 km will be overloaded by 2040 \cite{Barbar2021, CEA18}. Applying the flexible valuation framework to the representative feeders and scaling the total demand each feeder represents, we estimate that 29 and 140 GWh of battery storage NWA could be cost-effectively deployed across the four megacities in 2030 and 2040, respectively (see Table \ref{table:mega}). This would defer 15,914 km of network upgrades for 2030 and an additional 18,127 km for 2040. Table \ref{table:dls-results} highlights two option costs: (1) flexible budget through storage and deferred upgrades: total annualized cost of storage installed on the feeders and annualized deferred network upgrade costs (after battery storage NWA use is exhausted) and (2) the total annualized traditional network upgrade cost. Table \ref{table:dls-results} highlight the total budgets of the flexible valuation framework and traditional network upgrades across the four megacities. Total budget is calculated by summing the annualized investment cost of each year over a 30 years horizon. For traditional network investment, the total budget is the annualized investment cost of network upgrades times 30. For the flexible valuation framework, the total budget is the annualized investment cost of battery storage NWA for the deferral period and the annualized investment cost of deferred network investment for the rest of the planning horizon. Deploying battery storage NWA before traditional network upgrades produces capital cost savings of 16 and 15 \% in 2030 and 2040, respectively, on a total budget basis. More battery storage NWA is deployed per unit kilometer in 2040 than in 2030 due to the expected increase in the concentration of load during the peak hours as space cooling drives electricity demand growth \cite{Barbar2021}. Battery storage NWA is carried over from one stage to the next (stored in the value of $F$ of Eq. \ref{eqn:mdp}) existing system cost and remains available as long as it is dispatchable and network upgrades can be deferred. The resulting useful life of battery storage NWA ranges between 5 and 10 years.

\begin{table}[!ht]
\centering
\caption{Flexible valuation framework total results of megacity-level battery storage NWA analysis. Initial investment in 2020 with 10 year deferral period for periods 2030 and 2040. Distribution network line useful life is set to 30 years and battery storage useful life is set to 15 years.}
\begin{tabularx}{8cm}{l|cc}

\hline
Cost (in Millions of 2020 USD)&2030 &2040 \\
\hline
Annualized storage cost & \$207 & \$261 \\
Annualized deferred upgrades costs & \$76 & \$136  \\
\hline
Annualized traditional upgrades costs & \$117  & \$133 \\
\hline
\hline
Total flexible budget & \$2,932& \$5,324 \\
Total traditional budget & \$3,503 & \$6,266 \\
\hline
\end{tabularx}
\label{table:dls-results}
\end{table}

We evaluate the flexible valuation framework under the low and high cost storage scenarios (Table \ref{table:distro}). The result of battery storage NWA and deferred upgrades is the same in the low-cost and the mid-cost scenarios, which indicates that the binding constraint is dispatch --- i.e. the availability of off-peak network capacity throughout the day to charge the battery storage for peak hours discharge. This finding suggests that battery storage NWA may not be viable for the networks that are initially heavily loaded. Under the high storage cost scenario, we estimate that cost-effective battery storage NWA deployment would defer 11,752 km and 13,717 km of network upgrades in 2030 and 2040, respectively. Consequently, capital cost savings drop to 12\% and 10\% for the respective periods. Not surprisingly, the higher cost of storage implies less economic battery storage NWA. We further increase the cost of storage energy to 261 USD/kWh and power to 227 USD/kW and find that battery storage is eliminated and all overloaded lines are traditionally upgraded without any deferrals.

\begin{table}[!ht]
\centering
\caption{Storage cost impact on outputs of the flexible valuation framework applied to the four Indian megacities,  for year 2030. Low, mid and high storage capital cost assumptions are sourced from \citep{ATB2020}. Breakeven costs are 261 USD/kWh and 227 USD/kW for energy and power capacity respectively.}
\begin{tabular}{l|ccc}
\hline
 & Low& Mid& High\\
 \hline
Energy (USD/kWh) & 116 & 168 & 236  \\
Power (USD/kW) & 101 & 146 & 205  \\
\hline
Battery storage NWA (GWh) & 29 & 29 & 18 \\
Deferred lines (km) & 15,914 & 15,914 & 11,752\\
\hline
\end{tabular}
\label{table:distro}
\end{table}

As detailed earlier, battery storage NWA is driven by capital investment savings for utilities rather than the competitiveness of storage as a resource. The attractiveness of battery storage NWA has a proliferation potential in network-constrained environments where utilities have short-term financial commitments. Our results show that up to 29 GWh of battery storage capacity can serve as NWA to shift up to 7 GW of peak demand for a total of 140 hours in 2030 and up to 35 GW of peak demand for a total of 183 hours in 2040. This indicates that up to 338 and 741 GWh of peak electricity consumption can be shifted in 2030 and 2040 respectively. India's total electricity demand is projected to be 2.3 and 3.5 TWh with 347 and 626 GW of peak demand in 2030 and 2040 respectively, under the mid-range growth scenario \cite{Barbar2021}. If adopted at scale, the load-shifting potential of battery storage NWA can impact the dispatch of generators on the bulk-power system. Specifically, the battery must be charged in the day, and in EMDE where coal \cite{NREL2020} is the dominant baseload generation, the long-term cost and environmental benefits may not outweigh the short-term cost benefits at the distribution level.

\section{Discussion}\label{discussion}
Utilities in EMDE are primarily concerned with capital allocation owing to the high cost of financing. In the case of network equipment, we define the capital utilization rate (CUR) as the ratio of equipment loading in a given period $W_{t,y}$ to network capacity $M_t$. Based on the MDP simulation for mid-growth demand trajectory of the single network case in Delhi, we estimate a higher CUR for the storage and deferred network investment(59\%) as opposed to the case of traditional network investment (53\%). Additionally, utilities in EMDE face shorter-term financial commitments \cite{UN2021} due to a lack of long-term loan availability. Improving CUR will therefore serve utilities better to recover their investment and fulfill their financial commitments.

\section{Conclusion}\label{conclusion}

The flexible valuation framework presents an approach that combines system design optimization with multi-stage decision making under uncertainty for distribution network planning. The strategy assesses the feasibility of short-duration battery storage as an alternative to network upgrades given the uncertainty in demand growth. The simulation detailed for a single distribution network in the Delhi case shows that storage can shave the peak demand and thus prevent the network from overloading, which enables the deferral of the lumpy network upgrades to future periods when capital is cheaper (discounted) and uncertainty is lower. We compute the scaled-up effects of this strategy by applying it to the case of distribution networks across four Indian megacities, that result in an estimated 29 GWh and 140 GWh of storage capacity deployment in 2030 and 2040, respectively. We find that under reasonable cost assumptions for battery storage, high uncertainty of demand growth and high cost of capital, installing battery storage NWA and deferring traditional networks to the future is cost-effective. The flexible valuation framework enables utilities to adopt a \textit{wait and see} strategy with smaller initial investment costs when there is high uncertainty about future demand growth.

In our modeling, battery storage NWA is driven by capital savings in network investment planning, where we assume that the battery storage and the network lines are owned by the same entity (the utility). This may not always be the case and could result in different outcomes than presented here. The flexible valuation framework focuses only on the use of battery storage for network deferrals without considering the added value of storage in ancillary services and arbitrage (including selling energy back to the grid). On the other hand, the impact of peak-shifting battery storage on the bulk power system at transmission level is also not considered. Investigating the deployment of storage at the distribution level can be only looked at in isolation when the displacement of demand is a small fraction of the total demand. Further work to couple the distribution-level study with a transmission level is necessary to fully portray the value of battery storage NWA. Additionally, the system design optimization does not consider generation expansion on the distribution network to allow for DER proliferation (e.g., rooftop solar PV) due to the space constraints in the densely populated urban megacities of India where a building's rooftop space is minimal relative to its peak demand. Future work could consider investigating battery storage NWA with DERs on the distribution network which can be assessed using the same methodology presented in this paper by adding generation (\textit{a}) to multiple zones (\textit{z}) with the model described in section \ref{m2}. Finally, there are several input values, such as discount rate and demand growth, when modified, can impact the end result (and potentially lead to battery storage as an NWA being uneconomical). While Delhi shares similar traits with large EMDE cities (i.e. Cairo, Abuja), in terms of demand growth from space cooling and cost of capital, each environment is inherently different and must carefully be studied. The results demonstrate the value for battery storage NWA under certain circumstances in the Indian context and call for further investigation of such planning strategies in other EMDE countries.

\section{Acknowledgments}
This research was supported by MIT Energy Initiative's Low Carbon Energy Centers and the Future of Storage Study. We thank Tata Power Delhi Distribution Limited for their cooperation and contribution to the input data, the MIT Energy Initiative Future of Storage study team members for their review of the results, and Prof. Ignacio Perez-Arriaga and Jerome Nsengiyaremye for their feedback on the manuscript.

\appendix
\section{Optimization model}\label{optmodel}

The below model's nomenclature is presented in Table \ref{table:nomencl}.

\begin{eqnarray} \label{obj}
\begin{split}
\min & \sum_{a}\sum_{z}\big(\Omega^{size}_{a,z}\cdot (C^I_{a,z}+C^F_{a,z})\\
    &\qquad \qquad +\Omega^{energy}_{a,z}\cdot (C^e_{a,z}+C^{Fe}_{a,z})\cdot (1+C^d_{a,z})\\
    &\qquad \qquad +\Omega^{charge}_{a,z}\cdot (C^c_{a,z}+C^{Fc}_{a,z})\\
    &\qquad \qquad +\sum_t \pi_{a,t,z} \cdot (C^V_{a,z}+C^{Vf}_{a,z})\\
    &\qquad \qquad +\sum_t \psi_{a,t,z} \cdot (C^{Ve}_{a,z}+C^{Vc}_{a,z}) \\
    &\qquad \qquad +\sum_t n_{a,t,z} \cdot C^{start}_{a,z}\big)\\
    &+\sum_t\sum_z \chi_{t, z} \cdot C^\chi_{z,z'} \\
\end{split}
\end{eqnarray}

\begin{eqnarray} \label{dem}
\begin{split}
\textrm{s.t.}& \quad L_{t,  z} = \sum_a \pi_{a, t,  z} + \chi_{t,  z}  + \psi^{discharge}_{t,z}\\
&\qquad -\psi^{charge}_{t,z} + \phi_{t,z,z'}\\
&\forall z, z'\in Z \quad \forall t \in T
\end{split}
\end{eqnarray}

\begin{eqnarray} \label{vre}
\begin{split}
\pi_{a,t,z} \leq \Omega^{size}_{a,z} \cdot A_{a,t,z}  \\ 
\forall a \in R, \forall t \in T, \forall z \in Z
\end{split}
\end{eqnarray}

\begin{eqnarray}\label{stor1}
\begin{split}
    &\quad \Gamma_{a,t,z} = \Gamma_{a,t-1,z} - \frac{\psi^{discharge}_{a,t,z}}{\eta^{discharge}_{a,z}} + \eta^{charge}_{a,z} \cdot \psi^{charge}_{a, t,z}\\
    &\quad \forall a \in S,\forall t \in T^{interior} \in T, \forall z \in Z
\end{split}
\end{eqnarray}

\begin{eqnarray}\label{stor1_1}
\begin{split}
    &\quad \Gamma_{a,t,z} = \Gamma_{a,t^{period},z} - \frac{\psi^{discharge}_{a,t,z}}{\eta^{discharge}_{a,z}} + \eta^{charge}_{a,z} \cdot \psi^{charge}_{a, t,z}\\
    &\quad \forall a \in S, \forall t \in T^{start} , \forall z \in Z
\end{split}
\end{eqnarray}

\begin{eqnarray}\label{stor5}
\begin{split}
\Gamma_{a,t,z} \leq \delta_{a,z} \cdot \Omega^{energy}_{a,z} \\ \forall a \in S, \forall t \in T, \forall z \in Z
\end{split}
\end{eqnarray}

\begin{eqnarray}\label{stor3}
\begin{split}
\psi_{a,t,z}^{charge} \leq \Omega^{charge}_{a,z} \\ \forall a \in S, \forall t \in T, \forall z \in Z
\end{split}
\end{eqnarray}

\begin{eqnarray}\label{stor31}
\begin{split}
\psi_{a,t,z}^{charge} + \psi_{a,t,z}^{discharge} \leq \Omega^{charge}_{a,z} \\ \forall a \in S, \forall t \in T, \forall z \in Z
\end{split}
\end{eqnarray}

\begin{eqnarray}\label{stor4}
\begin{split}
\psi_{a,t,z} \leq \Gamma_{a,t-1,z} \\ \forall a \in S, \forall t \in T, \forall z \in Z
\end{split}
\end{eqnarray}

\begin{eqnarray}\label{therm1}
\begin{split}
v_{a,t,z} \leq \frac{\Omega^{size}_{a,z}}{\Omega^{unit}_{a,z}} \\ \forall a \in M, \forall t \in T, \forall z \in Z
\end{split}
\end{eqnarray}

\begin{eqnarray}\label{therm11}
\begin{split}
u_{a,t,z} \leq \frac{\Omega^{size}_{a,z}}{\Omega^{unit}_{a,z}} \\ \forall a \in M, \forall t \in T, \forall z \in Z
\end{split}
\end{eqnarray}

\begin{eqnarray}\label{therm12}
\begin{split}
n_{a,t,z} \leq \frac{\Omega^{size}_{a,z}}{\Omega^{unit}_{a,z}} \\ \forall a \in M, \forall t \in T, \forall z \in Z
\end{split}
\end{eqnarray}

\begin{eqnarray}\label{therm1_1}
\begin{split}
v_{a,t,z} = v_{a,t-1,z} + u_{a,t,z} + n_{a,t,z}\\ \forall a \in M, \forall t \in T^{interior}, \forall z \in Z
\end{split}
\end{eqnarray}

\begin{eqnarray}\label{therm1_2}
\begin{split}
v_{a,t,z} = v_{a,t+t^{period}-1,z} + u_{a,t,z} + n_{a,t,z}\\ \forall a \in M, \forall t \in T^{start}, \forall z \in Z
\end{split}
\end{eqnarray}

\begin{eqnarray}\label{therm2}
\begin{split}
\pi_{a,t,z} \geq \rho^{min}_{a,z} \cdot \Omega^{size}_{a,z} \cdot v_{a,t,z} \\ \forall a \in M, \forall t \in T, \forall z \in Z
\end{split}
\end{eqnarray}

\begin{eqnarray}\label{therm3}
\begin{split}
\pi_{a,t,z} \leq \rho^{max}_{a,z} \cdot \Omega^{size}_{a,z} \cdot v_{a,t,z} \\ \forall a \in M, \forall t \in T, \forall z \in Z
\end{split}
\end{eqnarray}

\begin{eqnarray}\label{therm4}
\begin{split}
\pi_{a,t,z} - \pi_{a,t-1,z} \leq \Omega^{size}_{a,z} \cdot \kappa_{a,z}^{up} \\ \forall a \in M, \forall t \in T, \forall z \in Z
\end{split}
\end{eqnarray}

\begin{eqnarray}\label{therm5}
\begin{split}
\pi_{a,t-1,z} - \pi_{a,t,z} \leq \Omega^{size}_{a,z} \cdot \kappa_{a,z}^{down} \\ \forall a \in M, \forall t \in T, \forall z \in Z
\end{split}
\end{eqnarray}

\begin{eqnarray}\label{net1}
\begin{split}
\phi_{t,z,z'} =  B_{z,z'} \cdot (\theta_{t,  z}-\theta_{t,z'}) \\ \forall z, z' \in Z, \forall t \in T
\end{split}
\end{eqnarray}

\begin{eqnarray}\label{net2}
\begin{split}
\phi_{t,z,z'} \leq \Phi^{max}_{z,z'}
\\ \forall i, j \in Z, \forall t \in T
\end{split}
\end{eqnarray}

\begin{eqnarray}\label{net3}
\begin{split}
\phi_{t,z,z'} \geq -\Phi^{max}_{z,z'}
\\ \forall i, j \in Z, \forall t \in T
\end{split}
\end{eqnarray}

\begin{eqnarray}\label{nnc}
\begin{split}
\Omega^{charge}_{a,z}, \Omega^{discharge}_{a,z}, \Omega^{size}_{a,z} \geq 0 \\
\pi_{a,t,z}, \chi_{t,z}, \Gamma_{a,t,z}  \geq 0 \\
\Psi^{charge}, \Psi^{discharge} \geq 0\\
\theta^{min}_{t,z,z'} \leq \theta_{t,z,z'} \leq \theta^{max}_{t,z,z'} \\
\Phi_{z,z'} \geq 0\\
\phi_{t} \quad \textrm{free}\\
\forall z,z' \in Z, \forall t \in T, \forall a \in M, R, S
\end{split}
\end{eqnarray}

\begin{table}[!ht]
\centering
\caption{Nomenclature of the electricity resource capacity expansion model of Appendix \ref{optmodel}}
\begin{tabular}{|c|c|}
\hline
\textbf{Set} & \textbf{Description} \\
\hline
$R$&Variable renewable energy resources\\
$S$&Battery storage resources\\
$M$&Thermal generation resources\\
$Z$&Power system zones\\
$T$&Hours of operation in a model period\\
$T^{interior}$&Interior time steps inside T\\
$T^{start}$&First time step of T\\
\hline
\textbf{Index} & \textbf{Description} \\
\hline
$a$&Generation resource\\
$t$&Time step\\
$z, z'$&Load zone\\
\hline
\textbf{Parameter} & \textbf{Description} \\
\hline
$t^{period}$&Total number of time steps\\
$w_t$&Time step weight\\
$C^I$&Investment cost (USD/MW)\\
$C^e$&Energy capacity investment cost (USD/MWh)\\
$C^c$&Charge investment cost (USD/MWh)\\
$C^d$&Battery energy capacity degradation per annum (\%)\\
$C^F$&Fixed operational cost (USD/MW-yr)\\
$C^{Fc}$&Fixed operational charge cost (USD/MWh-yr)\\
$C^V$&Variable cost (USD/MWh)\\
$C^{Vf}$&Fuel cost (USD/MWh)\\
$C^{Vc}$&Variable charge cost (USD/MWh)\\
$C^{\chi}$&Value of lost load (USD/MWh)\\
$C^\Phi$&Line upgrade cost (USD/MW)\\
$C^{start}$&Startup cost (USD/MW)\\
$\mu$&Storage round-trip efficiency\\
$A$&Generation availability profile\\
$\Omega^{max}$&Maximum generation capacity\\
$\rho^{min}$&Minimum generation power\\
$\rho^{max}$&Maximum generation power\\
$\kappa^{up}$&Ramp up limit\\
$\kappa^{down}$&Ramp down limit\\
$B$&Line susceptance\\
$\Phi^{max}$&Maximum line power capacity\\
$\delta$&Storage depth of discharge\\
$\eta$&Storage efficiency\\
$\Omega^{unit}$&Generation unit capacity\\
$\theta^{min}, \theta^{max}$&Minimum and maximum voltage angle\\

\hline
\textbf{Variable} & \textbf{Description} \\
\hline

$\Omega^{size}_{a,z}$&Generation capacity\\
$\Omega^{energy}_{a,z}$&Energy capacity\\
$\Omega^{charge}_{a,z}$&Charge capacity\\
$\pi_{a,t,z}$&Power output\\
$v_{a,t,z}$&Number of units committed\\
$u_{a,t,z}$&Number of startup decisions\\
$n_{a,t,z}$&Number of shutdown decisions\\
$\Psi^{charge}_{a,t,z}$&Storage charge\\
$\Psi^{discharge}_{a,t,z}$&Storage discharge\\
$\chi_{t,z}$&Non-served energy\\
$\Gamma_{a,t,z}$&Storage state of charge\\
$\theta_{t,z}$&Line voltage angle\\
$\phi_{t,z,z'}$&Line power flow\\
\hline
\end{tabular}
\label{table:nomencl}
\end{table}

\bibliographystyle{elsarticle-num}

\bibliography{main}

\begin{thebibliography}{10}
\expandafter\ifx\csname url\endcsname\relax
  \def\url#1{\texttt{#1}}\fi
\expandafter\ifx\csname urlprefix\endcsname\relax\def\urlprefix{URL }\fi
\expandafter\ifx\csname href\endcsname\relax
  \def\href#1#2{#2} \def\path#1{#1}\fi

\bibitem{UoF2016}
I.~P\'erez-Arriaga,
  \href{https://energy.mit.edu/wp-content/uploads/2016/12/Utility-of-the-Future-Full-Report.pdf}{{Utility
  of the future: An MIT Energy Initiative response to an industry in
  transition}}, Tech. rep., Massachusetts Institute of Technology (2016).
\newline\urlprefix\url{https://energy.mit.edu/wp-content/uploads/2016/12/Utility-of-the-Future-Full-Report.pdf}

\bibitem{Barbar2019}
M.~Barbar, \href{https://dspace.mit.edu/handle/1721.1/122752}{{Resiliency and
  reliability planning of the electric grid in natural disaster affected
  areas}}, Master's thesis, Massachusetts Institute of Technology (2019).
\newline\urlprefix\url{https://dspace.mit.edu/handle/1721.1/122752}

\bibitem{Rao2019}
N.~D. Rao, J.~Min, A.~Mastrucci,
  \href{https://doi.org/10.1038/s41560-019-0497-9}{{Energy requirements for
  decent living in India, Brazil and South Africa}}, Nature Energy 4~(12)
  (2019) 1025--1032.
\newblock \href {https://doi.org/10.1038/s41560-019-0497-9}
  {\path{doi:10.1038/s41560-019-0497-9}}.
\newline\urlprefix\url{https://doi.org/10.1038/s41560-019-0497-9}

\bibitem{su10051440}
K.~Olaniyan, B.~C. McLellan, S.~Ogata, T.~Tezuka,
  \href{https://www.mdpi.com/2071-1050/10/5/1440}{Estimating residential
  electricity consumption in {N}igeria to support energy transitions},
  Sustainability 10~(5) (2018).
\newblock \href {https://doi.org/10.3390/su10051440}
  {\path{doi:10.3390/su10051440}}.
\newline\urlprefix\url{https://www.mdpi.com/2071-1050/10/5/1440}

\bibitem{Barbar2021}
M.~Barbar, D.~S. Mallapragada, M.~Alsup, R.~Stoner,
  \href{https://doi.org/10.1038/s41597-021-00951-6}{Scenarios of future
  {I}ndian electricity demand accounting for space cooling and electric vehicle
  adoption}, Scientific Data 8~(178) (Jul. 2021).
\newblock \href {https://doi.org/10.1038/s41597-021-00951-6}
  {\path{doi:10.1038/s41597-021-00951-6}}.
\newline\urlprefix\url{https://doi.org/10.1038/s41597-021-00951-6}

\bibitem{McNeil2013}
M.~A. McNeil, V.~E. Letschert, S.~de~la Rue~du Can, J.~Ke,
  \href{https://doi.org/10.1007/s12053-012-9182-6}{Bottom--up energy analysis
  system (buenas)---an international appliance efficiency policy tool}, Energy
  Efficiency 6~(2) (2013) 191--217.
\newblock \href {https://doi.org/10.1007/s12053-012-9182-6}
  {\path{doi:10.1007/s12053-012-9182-6}}.
\newline\urlprefix\url{https://doi.org/10.1007/s12053-012-9182-6}

\bibitem{IEAcooling18}
{International Energy Agency (IEA)}, The future of cooling - opportunities for
  energy efficient air conditioning, Tech. rep., IEA (2018).

\bibitem{UN2021}
{Economic and Social Council}, {Short-Term Financing, Creation of Repo Market
  Crucial to Assist Poor Countries Facing Escalating Debt, Economic
  Contraction, Speakers Tell Financing for Development Forum} (2021).

\bibitem{WEC2009}
{World Energy Council}, Transmission and distribution in {I}ndia (2009).

\bibitem{WB2014}
S.~Pargal, S.~G. Banerjee, More Power to {I}ndia: The Challenge of Electricity
  Distribution, World Bank, Washington, DC, 2014, license: Creative Commons
  Attribution CC BY 3.0 IGO.
\newblock \href {https://doi.org/10.1596/978-1-4648-0233-1}
  {\path{doi:10.1596/978-1-4648-0233-1}}.

\bibitem{brookings}
S.~Ali,
  \href{https://www.brookings.edu/wp-content/uploads/2018/10/The-future-of-Indian-electricity-demand.pdf}{The
  future of {I}ndian electricity demand: How much, by whom and under what
  conditions?}, Tech. rep., The Brookings Institute (2018).
\newline\urlprefix\url{https://www.brookings.edu/wp-content/uploads/2018/10/The-future-of-Indian-electricity-demand.pdf}

\bibitem{IndiaEnergyOutlook}
{International Energy Agency},
  \href{https://www.iea.org/reports/india-energy-outlook-2021}{India Energy
  Outlook 2021}, IEA, 2021.
\newline\urlprefix\url{https://www.iea.org/reports/india-energy-outlook-2021}

\bibitem{CEA2018}
{Central Electricity Authority}, Growth of electricity sector in {I}ndia from
  1947-2019 (2018).

\bibitem{Evans2020}
A.~Evans, \href{http://dspace.mit.edu/handle/1721.1/7582}{{The value of
  flexibility : application of real options analysis to electricity network
  investments}}, Master's thesis, Massachusetts Institute of Technology (2020).
\newline\urlprefix\url{http://dspace.mit.edu/handle/1721.1/7582}

\bibitem{ifc2019}
{International Finance Council Corporation}, The dirty footprint of the broken
  grid: The impacts of fossil fuel back-up generators in developing countries,
  Tech. rep., World Bank Group (2019).

\bibitem{ATB2020}
{National Renewable Energy Laboratory (NREL)}, {Annual Technology Baseline:
  Electricity} (2020).

\bibitem{6344193}
A.~Q. Jakhrani, A.~R.~H. Rigit, A.-K. Othman, S.~R. Samo, S.~A. Kamboh,
  Estimation of carbon footprints from diesel generator emissions, in: 2012
  International Conference on Green and Ubiquitous Technology, 2012, pp.
  78--81.
\newblock \href {https://doi.org/10.1109/GUT.2012.6344193}
  {\path{doi:10.1109/GUT.2012.6344193}}.

\bibitem{6450147}
M.~E. {Samper}, A.~{Vargas}, Investment decisions in distribution networks
  under uncertainty with distributed generation—part i: Model formulation,
  IEEE Transactions on Power Systems 28~(3) (2013) 2331--2340.
\newblock \href {https://doi.org/10.1109/TPWRS.2013.2239666}
  {\path{doi:10.1109/TPWRS.2013.2239666}}.

\bibitem{KIM2017918}
K.~Kim, H.~Park, H.~Kim,
  \href{https://www.sciencedirect.com/science/article/pii/S1364032116307493}{Real
  options analysis for renewable energy investment decisions in developing
  countries}, Renewable and Sustainable Energy Reviews 75 (2017) 918--926.
\newblock \href {https://doi.org/https://doi.org/10.1016/j.rser.2016.11.073}
  {\path{doi:https://doi.org/10.1016/j.rser.2016.11.073}}.
\newline\urlprefix\url{https://www.sciencedirect.com/science/article/pii/S1364032116307493}

\bibitem{DAS2020100482}
P.~Das, P.~Mathuria, R.~Bhakar, J.~Mathur, A.~Kanudia, A.~Singh,
  \href{https://www.sciencedirect.com/science/article/pii/S2211467X20300353}{Flexibility
  requirement for large-scale renewable energy integration in {I}ndian power
  system: Technology, policy and modeling options}, Energy Strategy Reviews 29
  (2020) 100482.
\newblock \href {https://doi.org/https://doi.org/10.1016/j.esr.2020.100482}
  {\path{doi:https://doi.org/10.1016/j.esr.2020.100482}}.
\newline\urlprefix\url{https://www.sciencedirect.com/science/article/pii/S2211467X20300353}

\bibitem{GRIMM2020114017}
V.~Grimm, J.~Grübel, B.~Rückel, C.~Sölch, G.~Zöttl,
  \href{https://www.sciencedirect.com/science/article/pii/S0306261919317040}{Storage
  investment and network expansion in distribution networks: The impact of
  regulatory frameworks}, Applied Energy 262 (2020) 114017.
\newblock \href
  {https://doi.org/https://doi.org/10.1016/j.apenergy.2019.114017}
  {\path{doi:https://doi.org/10.1016/j.apenergy.2019.114017}}.
\newline\urlprefix\url{https://www.sciencedirect.com/science/article/pii/S0306261919317040}

\bibitem{PILPOLA2019100368}
S.~Pilpola, P.~D. Lund,
  \href{https://www.sciencedirect.com/science/article/pii/S2211467X19300550}{Different
  flexibility options for better system integration of wind power}, Energy
  Strategy Reviews 26 (2019) 100368.
\newblock \href {https://doi.org/https://doi.org/10.1016/j.esr.2019.100368}
  {\path{doi:https://doi.org/10.1016/j.esr.2019.100368}}.
\newline\urlprefix\url{https://www.sciencedirect.com/science/article/pii/S2211467X19300550}

\bibitem{SGOURIDIS20138}
S.~Sgouridis, S.~Griffiths, S.~Kennedy, A.~Khalid, N.~Zurita,
  \href{https://www.sciencedirect.com/science/article/pii/S2211467X13000333}{A
  sustainable energy transition strategy for the {United Arab Emirates}:
  Evaluation of options using an integrated energy model}, Energy Strategy
  Reviews 2~(1) (2013) 8--18, strategy Options and Models for the Middle East
  and North Africa (MENA) Energy Transition.
\newblock \href {https://doi.org/https://doi.org/10.1016/j.esr.2013.03.002}
  {\path{doi:https://doi.org/10.1016/j.esr.2013.03.002}}.
\newline\urlprefix\url{https://www.sciencedirect.com/science/article/pii/S2211467X13000333}

\bibitem{Jenkins2017}
J.~Jenkins, N.~Sepulveda,
  \href{https://energy.mit.edu/wp-content/uploads/2017/10/Enhanced-Decision-Support-for-a-Changing-Electricity-Landscape.pdf}{{Enhanced
  Decision Support for a Changing Electricity Landscape: the GenX Configurable
  Electricity Resource Capacity Expansion Model}}, Tech. rep., MIT Energy
  Initiative (2017).
\newline\urlprefix\url{https://energy.mit.edu/wp-content/uploads/2017/10/Enhanced-Decision-Support-for-a-Changing-Electricity-Landscape.pdf}

\bibitem{Knueven2020}
B.~Knueven, J.~Ostrowski, J.-P. Watson,
  \href{https://doi.org/10.1287/ijoc.2019.0944}{On mixed-integer programming
  formulations for the unit commitment problem}, {INFORMS} Journal on Computing
  (Jun. 2020).
\newblock \href {https://doi.org/10.1287/ijoc.2019.0944}
  {\path{doi:10.1287/ijoc.2019.0944}}.
\newline\urlprefix\url{https://doi.org/10.1287/ijoc.2019.0944}

\bibitem{SAHLBERG2021100714}
A.~Sahlberg, B.~Khavari, A.~Korkovelos, F.~{Fuso Nerini}, M.~Howells,
  \href{https://www.sciencedirect.com/science/article/pii/S2211467X21001000}{A
  scenario discovery approach to least-cost electrification modelling in
  {Burkina Faso}}, Energy Strategy Reviews 38 (2021) 100714.
\newblock \href {https://doi.org/https://doi.org/10.1016/j.esr.2021.100714}
  {\path{doi:https://doi.org/10.1016/j.esr.2021.100714}}.
\newline\urlprefix\url{https://www.sciencedirect.com/science/article/pii/S2211467X21001000}

\bibitem{OLSSON2021100705}
J.~M. Olsson, F.~Gardumi,
  \href{https://www.sciencedirect.com/science/article/pii/S2211467X21000912}{Modelling
  least cost electricity system scenarios for {B}angladesh using osemosys},
  Energy Strategy Reviews 38 (2021) 100705.
\newblock \href {https://doi.org/https://doi.org/10.1016/j.esr.2021.100705}
  {\path{doi:https://doi.org/10.1016/j.esr.2021.100705}}.
\newline\urlprefix\url{https://www.sciencedirect.com/science/article/pii/S2211467X21000912}

\bibitem{9347026}
P.~Shi, Y.~Li, Y.~Du, Q.~Cai, H.~Chen, Distribution network planning
  considering uncertainty of incremental distribution network access, in: 2020
  IEEE 4th Conference on Energy Internet and Energy System Integration (EI2),
  2020, pp. 2244--2247.
\newblock \href {https://doi.org/10.1109/EI250167.2020.9347026}
  {\path{doi:10.1109/EI250167.2020.9347026}}.

\bibitem{9088223}
H.~Wang, Z.~Yan, M.~Shahidehpour, Q.~Zhou, X.~Xu, Optimal energy storage
  allocation for mitigating the unbalance in active distribution network via
  uncertainty quantification, IEEE Transactions on Sustainable Energy 12~(1)
  (2021) 303--313.
\newblock \href {https://doi.org/10.1109/TSTE.2020.2992960}
  {\path{doi:10.1109/TSTE.2020.2992960}}.

\bibitem{batteries6040056}
P.~Boonluk, A.~Siritaratiwat, P.~Fuangfoo, S.~Khunkitti,
  \href{https://www.mdpi.com/2313-0105/6/4/56}{Optimal siting and sizing of
  battery energy storage systems for distribution network of distribution
  system operators}, Batteries 6~(4) (2020).
\newblock \href {https://doi.org/10.3390/batteries6040056}
  {\path{doi:10.3390/batteries6040056}}.
\newline\urlprefix\url{https://www.mdpi.com/2313-0105/6/4/56}

\bibitem{9582899}
M.~Jooshaki, S.~Fattaheian-Dehkordi, M.~Fotuhi-Firuzabad, M.~Lehtonen, Planning
  a flexible distribution network with energy storage systems considering the
  uncertainty of renewable sources and demand, in: CIRED 2020 Berlin Workshop
  (CIRED 2020), Vol. 2020, 2020, pp. 132--135.
\newblock \href {https://doi.org/10.1049/oap-cired.2021.0288}
  {\path{doi:10.1049/oap-cired.2021.0288}}.

\bibitem{7353222}
G.~Muñoz-Delgado, J.~Contreras, J.~M. Arroyo, Multistage generation and
  network expansion planning in distribution systems considering uncertainty
  and reliability, IEEE Transactions on Power Systems 31~(5) (2016) 3715--3728.
\newblock \href {https://doi.org/10.1109/TPWRS.2015.2503604}
  {\path{doi:10.1109/TPWRS.2015.2503604}}.

\bibitem{gitgenx}
N.~Sepulveda, J.~Jenkins, D.~Mallapragada, A.~Schwartz, N.~Patankar, Q.~Xu,
  J.~Morri, S.~Chakrabarti, \href{https://github.com/GenXProject/GenX}{Source
  code for: Genx}, \emph{Github}. \url{https://github.com/GenXProject/GenX}
  (2021).
\newline\urlprefix\url{https://github.com/GenXProject/GenX}

\bibitem{gitegret}
B.~Knueven, J.~Ostrowski, J.-P. Watson,
  \href{https://github.com/grid-parity-exchange/Egret}{Source code for: Egret},
  \emph{Github}. \url{https://github.com/grid-parity-exchange/Egret} (2021).
\newline\urlprefix\url{https://github.com/grid-parity-exchange/Egret}

\bibitem{tata_tariff}
{Tata Power Delhi Distribution Limited}, Customer solutions tariff details and
  net-metering (2020).

\bibitem{hart2017pyomo}
W.~E. Hart, C.~D. Laird, J.-P. Watson, D.~L. Woodruff, G.~A. Hackebeil, B.~L.
  Nicholson, J.~D. Siirola, Pyomo--optimization modeling in python, 2nd
  Edition, Vol.~67, Springer Science \& Business Media, 2017.

\bibitem{cplex2009v12}
I.~I. Cplex, V12. 1: User’s manual for cplex, International Business Machines
  Corporation 46~(53) (2009) 157.

\bibitem{gompertz}
C.~P. Winsor, \href{http://www.jstor.org/stable/86156}{The gompertz curve as a
  growth curve}, Proceedings of the National Academy of Sciences of the United
  States of America 18~(1) (1932) 1--8.
\newline\urlprefix\url{http://www.jstor.org/stable/86156}

\bibitem{WB2021}
{World Bank Group}, Electric power consumption (kwh per capita), \url{
  https://data.worldbank.org/} (2020).

\bibitem{1709447}
K.~{Rudion}, A.~{Orths}, Z.~A. {Styczynski}, K.~{Strunz}, Design of benchmark
  of medium voltage distribution network for investigation of dg integration,
  in: 2006 IEEE Power Engineering Society General Meeting, 2006, pp. 6 pp.--.
\newblock \href {https://doi.org/10.1109/PES.2006.1709447}
  {\path{doi:10.1109/PES.2006.1709447}}.

\bibitem{tata}
{Tata Power Delhi Distribution Limited}, {Technical specification cover sheets:
  33KV and 11KV Grid} (2012).

\bibitem{MOCI2019}
{Office of the economic adviser department for promotion of industry and
  internal trade. Ministry of Commerce and Industry. Government of India.},
  \href{https://eaindustry.nic.in/}{Index files for wpi series} (2019).
\newline\urlprefix\url{https://eaindustry.nic.in/}

\bibitem{NRELCD2019}
K.~Horowitz, 2019 distribution system upgrade unit cost database current
  version, \url{ https://data.nrel.gov/submissions/101} (2019).

\bibitem{LAZARD2020}
{Lazard Ltd.}, Lazard’s levelized cost of storage analysis --- version 6.0
  (2020).

\bibitem{prayas2012}
S.~Nhalur, A.~Josey,
  \href{https://www.prayaspune.org/peg/publications/item/176-electricity-in-megacities.html}{Electricity
  in megacities}, Tech. rep., Prayas Energy Group (2012).
\newline\urlprefix\url{https://www.prayaspune.org/peg/publications/item/176-electricity-in-megacities.html}

\bibitem{rbsa}
{RBSA Advisors},
  \href{https://rbsa.in/cost-of-capital-in-india-4th-edition/}{Cost of capital
  in {I}ndia} (2020).
\newline\urlprefix\url{https://rbsa.in/cost-of-capital-in-india-4th-edition/}

\bibitem{CEA18}
{Central Electricity Authority}, Guidelines for distribution utilities for
  development of distribution infrastructure (2018).

\bibitem{TPDDL}
{Tata Power Delhi Distribution Limited},
  \href{https://www.tatapower.com/investor-relations/annual-reports-archive.aspx}{Excellence
  journey progress report} (2016).
\newline\urlprefix\url{https://www.tatapower.com/investor-relations/annual-reports-archive.aspx}

\bibitem{NREL2020}
A.~Rose, l.~Chernyakhovskiy, D.~Palchak, S.~Koebrich, M.~Joshi,
  \href{https://www.nrel.gov/docs/fy20osti/76153.pdf}{{Least-Cost Pathways for
  India’s Electric Power Sector}}, Tech. rep., National Renewable Energy
  Laboratory, Golden, CO (2020).
\newline\urlprefix\url{https://www.nrel.gov/docs/fy20osti/76153.pdf}

\end{thebibliography}


\end{document}